\documentclass[a4paper,aps,prl,showpacs,twocolumn,superscriptaddress]{revtex4-1} 

\usepackage[utf8]{inputenc}  
\usepackage[british]{babel}  
\usepackage{graphicx} 
\usepackage{tikz}
\usetikzlibrary{arrows}
\usepackage{amsmath,dsfont}
\usepackage{amssymb}
\usepackage{hyperref}
\usepackage{graphicx}  
\usepackage[caption=false]{subfig}
\usepackage{dcolumn}   
\usepackage{bm}        
\usepackage{xr}
\usepackage{xcite}
\externalcitedocument{hardy-ghz-supp}

\newcommand\ue{\mathrel{\bullet\mkern-3mu{-}\mkern-3mu\bullet}}

\usepackage{xcolor}

\newcommand{\ket}[1]{\ensuremath{|#1\rangle}}

\newcommand{\bra}[1]{\ensuremath{\langle#1|}}

\newcommand{\braket}[2]{\ensuremath{\langle #1|#2\rangle}}
\newcommand{\ketbra}[1]{\ensuremath{| #1 \rangle \!\langle #1 |}}

\newcommand{\id}{\ensuremath{\mathds{1}}}

\newcommand{\qed}{\ensuremath{\hfill \Box}}

\newcommand{\ba}{\begin{eqnarray}}
\newcommand{\ea}{\end{eqnarray}}
\newcommand{\ban}{\begin{eqnarray*}}
\newcommand{\ean}{\end{eqnarray*}}
\newcommand{\be}{\begin{equation}}
\newcommand{\ee}{\end{equation}}
\newcommand*\cn[1]{\tikz[baseline=(char.base)]{
\node[shape=circle,draw,inner sep=0.1pt,line width=0.08mm, minimum size=0.3cm] (char) {{\fontsize{0.22cm}{0.25cm}\selectfont $#1$}};}}
\makeatletter
\DeclareRobustCommand{\cev}[1]{%
  {\mathpalette\do@cev{#1}}%
}
\newcommand{\do@cev}[2]{%
  \vbox{\offinterlineskip
    \sbox\z@{$\m@th#1 x$}%
    \ialign{##\cr
      \hidewidth\reflectbox{$\m@th#1\vec{}\mkern4mu$}\hidewidth\cr
      \noalign{\kern-\ht\z@}
      $\m@th#1#2$\cr
    }%
  }%
}
\makeatother

\begin{document}
\nonfrenchspacing
\title{Proof of the Peres conjecture for contextuality}

\author{Zhen-Peng Xu}
\email{Zhen-Peng.Xu@uni-siegen.de}
\affiliation{Naturwissenschaftlich-Technische 
Fakult\"{a}t, Universit\"{a}t Siegen, Walter-Flex-Straße 3, 57068 Siegen, Germany}
\author{Jing-Ling Chen}
\email{chenjl@nankai.edu.cn}
\affiliation{Theoretical Physics Division, Chern Institute of Mathematics,
Nankai University, Tianjin 300071, China}
\author{Otfried G\"{u}hne}
\email{otfried.guehne@uni-siegen.de}
\affiliation{Naturwissenschaftlich-Technische 
Fakult\"{a}t, Universit\"{a}t Siegen, Walter-Flex-Straße 3, 57068 Siegen, Germany}

\date{\today}  


\begin{abstract}
A central result in the foundations of quantum mechanics is the 
Kochen-Specker theorem. In short, it states that quantum mechanics 
cannot be reconciled with classical models that are noncontextual 
for ideal measurements. The first explicit derivation by 
Kochen and Specker was rather complex, but considerable 
simplifications have been achieved thereafter. We propose 
a systematic approach to find minimal Hardy-type and 
Greenberger-Horne-Zeilinger-type (GHZ-type) proofs 
of the Kochen-Specker theorem, these are characterized 
by the fact that the predictions of classical models are 
opposite to the predictions of quantum mechanics. 
Based on our results, we show that the Kochen-Specker set with $18$ 
vectors from Cabello {\it et al.} [A. Cabello {\it et al.}, Phys. Lett. 
A {\bf 212}, 183 (1996)] is the minimal set for any dimension, verifying 
a long-standing conjecture by Peres. Our results allow to identify minimal 
contextuality scenarios and to study their usefulness for 
information processing.
\end{abstract}
\maketitle

\definecolor{bgcolor}{RGB}{46,52,64}
\definecolor{ttcolor}{RGB}{214,220,231}

{\it Introduction.---}
The fact that quantum mechanics cannot be described by 
noncontextual hidden variable models is known as the 
Kochen-Specker (KS) theorem~\cite{kochen1967problem}. Since the derivation of 
experimentally testable inequalities \cite{cabello2008experimentally, 
klyachko2008simple}, plenty of theoretical~\cite{yu2012stateindependent, 
cabello2014graph, howard2014contextuality, KunjwalPRL2015}
and experimental \cite{Kirchmair2009, Amselem2009, Moussa2010, Zhang2013, Mazurek2016} 
works have been carried out. Consequently, contextuality 
has been linked to tasks in quantum computation \cite{howard2014contextuality}
and randomness generation \cite{miller17}, see also \cite{budroni2020} for a 
review.

Bell nonlocality can be seen as a form of contextuality, where
the locality assumption on separated observers enforces the 
measurements to behave in a non-contextual manner. Bell's theorem 
was originally proved by the derivation of Bell inequalities 
\cite{bell1966problem}, which hold for classical
theories, but are violated in quantum mechanics. After that, however, 
other proofs have emerged. For instance, Hardy's proof~\cite{hardy1992quantum} 
of nonlocality without inequalities is sometimes considered to be of 
``the simplest form"~\cite{mermin1995best}. In Hardy's proof, one shows
that certain events can never happen in a classical hidden variable model, 
while they can happen with a non-vanishing probability in the quantum case. 
An even stronger version is due to Greenberger, Horne and Zeilinger 
(GHZ)~\cite{greenberger1989going}, where some events are excluded in a 
classical model, but they occur {\it with certainty}
in the quantum case. Such examples are not only considered for mathematical
clarity or beauty; in practice they give strong tests for various forms
of multiparticle entanglement \cite{PhysRevLett.112.140404, Abramsky_2016, jiang2018generalized} 
and the resulting arguments 
are relevant for information processing. For instance, it has been shown that 
the building block for the original GHZ argument can be seen as a resource, 
enabling new computations. More precisely, GHZ correlations are the minimal 
resource to promote the parity computer to classical universality \cite{anders2009computational}.

A key point of the Kochen-Specker theorem is that its proof can 
be viewed as a finite version Gleason's theorem~\cite{gleason1957measures}, 
in the sense that it shows that already for a finite collection of 
measurements, dispersion-free states are impossible. This raises the 
question for the simplest possible proof. The original proof 
by Kochen and Specker (KS)~\cite{kochen1967problem} is a proof 
by contradiction, where a set of $117$ vectors in a three-dimensional 
space was used. If viewed as measurements in quantum mechanics, these
vectors obey certain conditions of mutual exclusivity, which cannot be
reproduced in a classical model; a set of vectors with this property
is also called a KS set. In the last 30 years, a march to find 
simpler KS proofs has been carried out~\cite{peres1991two, 
kernaghan1995kochen,cabello1996bell}, see also \cite{adanijqi} for 
an overview. In this respect, the proof of Cabello, Estebaranz, and 
Garc\'{i}a-Alcaine (CEG) using $18$ vectors in a four-dimensional space 
is the simplest one known, and is often used as {\it the} basic explanation 
of theorem \cite{stanford}. It was then conjectured 
by Peres~\cite{peres2003what} that this is indeed the optimal one, but 
the proof of this conjecture is still missing \footnote{The paper \cite{adanijqi} 
suggests a proof strategy, but it does not work for all dimensions [A. Cabello, 
private communication].}.
Besides these works, which follow the logic of the original article 
and construct KS sets, also Hardy-type proofs and GHZ-type proofs have 
been introduced~\cite{cabello2013simple,cabello1996bell}, and minimal 
inequalities for state-independent contextuality have been derived 
\cite{yu2012stateindependent,yu2015proof, kleinmannprl}.

\begin{table*}[t!!]
\centering
\begin{tabular}{|r|r|r|r|r|r|r|r|r|}
		\hline\hline
		$C_1\quad$  &  $C_2\quad$ &  $C_3\quad$ &  $C_4\quad$ &  $C_5\quad$ &  $C_6\quad$ &  $C_7\quad$ &  $C_8\quad$ &  $C_9\quad$\\ \hline\rule{0pt}{2.4ex}
                $\cn{0}:1000$ & $\cn{3}:01\bar{1}0$ & $\cn{6}:\bar{1}111$ &  $\cn{9}:010\bar{1}$ & $\cn{12}:1111$ &  $\cn{15}:001\bar{1}$ & $\cn{17}:0100$ & $\cn{2}:0110$ & $\cn{5}:111\bar{1}$\\  
                $\cn{1}:0001$ & $\cn{4}:1001$ &  $\cn{7}:11\bar{1}1$ &  $\cn{10}:10\bar{1}0$ & $\cn{13}:11\bar{1}\bar{1}$ & $\cn{16}:0011$ & $\cn{1}:0001$ & $\cn{4}:1001$ &  $\cn{16}:0011$\\
                $\cn{2}:0110$ & $\cn{5}:111\bar{1}$ &  $\cn{8}:1010$ &  $\cn{11}:1\bar{1}1\bar{1}$ & $\cn{14}:1\bar{1}00$ & $\cn{17}:0100$ &  $\cn{8}:1010$ & $\cn{11}:1\bar{1}1\bar{1}$ &  $\cn{14}:1\bar{1}00$\\
                $\cn{3}:01\bar{1}0$ & $\cn{6}:\bar{1}111$ & $\cn{9}:010\bar{1}$ & $\cn{12}:1111$ & $\cn{15}:001\bar{1}$ & $\cn{0}:1000$ &  $\cn{10}:10\bar{1}0$ &  $\cn{13}:11\bar{1}\bar{1}$ &  $\cn{7}:11\bar{1}1$\\
		\hline\hline
\end{tabular}
\caption{The Kochen-Specker set of vectors from Cabello and coworkers 
\cite{cabello1996bell}. This consists of 18 vectors in four-dimensional 
space, where  $\textcircled{\fontsize{0.22cm}{0.25cm}\selectfont {k}}:abcd$ 
means  $\ket{\psi_k} = (a,b,c,d)$ and $\bar{1} :=-1$, the normalization
is dropped. These vectors form $9$ complete contexts $C_i$, written as columns
in the table. A non-contextual model must, for each hidden variable, assign
to exactly one vector in each context the value $1$. So, in total the assignment
to $9$ entries in the table  must be $1$. On the other hand,  
since each vector $\textcircled{\fontsize{0.22cm}{0.25cm}\selectfont {k}}$ appears 
twice (e.g., $\textcircled{\fontsize{0.22cm}{0.25cm}\selectfont {14}}$ 
appears in $C_5$ and $C_9$), the value $1$ must be assigned to an even number 
of entries in the table, if the assignment is independent of the context. 
Thus, one has a contradiction and such an assignment is not possible.}\label{tab:context}
\end{table*}

In this paper we present a systematic method based on graph 
theory to construct Hardy-type and GHZ-type arguments for 
quantum contextuality with the minimal number of measurements. 
With this, we find the minimal GHZ-type proof, which needs 
$10$ measurement events. The key observation of our approach
is a connection between KS sets of vectors and GHZ-type proofs 
that use a subset of the vectors. Using this, we then can go 
on and show that $18$-vector proof by CEG is indeed the optimal 
one in any dimension, putting the Peres conjecture at rest. 
The minimal GHZ-type proof and the $18$-vector proof by CEG are 
therefore the basic building blocks of contextuality and the key 
to understand the role of it in information processing. We add 
that the GHZ-type argument with 10 events was noted before 
\cite{cabello1996bell} and similar graph-theoretical calculations
were carried out \cite{amselem2012experimental}, but only the
connection between GHZ-type proofs and KS sets allows to draw
far-ranging consequences for the Peres conjecture.

{\it Kochen-Specker sets of vectors.---}
First, recall the notion of ideal (i.e., repeatable and minimally disturbing) measurements and 
{\it events}, which are a combination of an ideal measurement and one 
of its outcomes. In quantum theory, ideal events are represented by 
vectors, corresponding to a measurement outcome. Two events are 
said to be {\it exclusive} if, for any state, one event cannot happen 
if the other happens. In the quantum mechanical description, two orthogonal
vectors represent a pair of exclusive events. A {\it context} 
is a set of compatible (i.e., non-disturbing) measurements, building a set 
of mutually exclusive events, and a context is said to be {\it complete} 
if always one event happens, whatever the input state is. For instance, 
in quantum mechanics the eigenvectors of an observable 
form a complete context, they can all be measured at the same time 
and exactly one result occurs. 

Given these notions, one can ask whether the quantum mechanical predictions
can be reproduced by a classical hidden variable model. For a given hidden
variable, such a model needs to assign to any ideal event the values
$1$ or $0$, depending on whether the event takes place or not. Here, it is
natural to make the constraint of non-contextuality: The value assigned to an event should
be independent of the contexts where the event belongs to. Such a
model is called a noncontextual hidden variable (NCHV) model. 

\begin{figure}[t]
		\centering
		\includegraphics[width=0.26\textwidth]{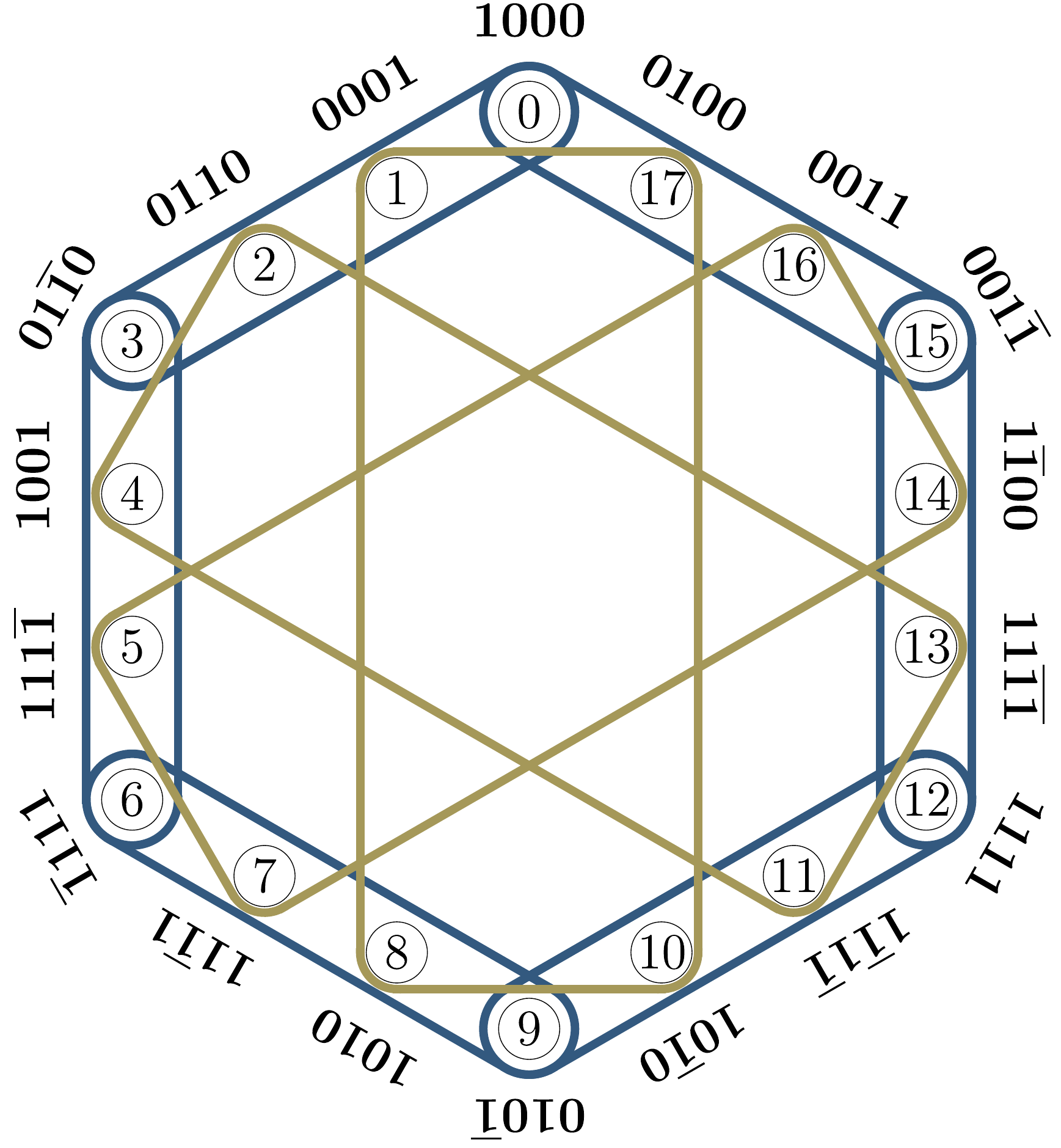}
\caption{The original CEG vectors from Table I 
can be arranged into  $9$ complete contexts.
Each complete context is represented by a closed curve, but incomplete 
contexts like 
$\{\textcircled{\fontsize{0.22cm}{0.25cm}\selectfont {0}},
\textcircled{\fontsize{0.22cm}{0.25cm}\selectfont {9}}\}$
or
$\{\textcircled{\fontsize{0.22cm}{0.25cm}\selectfont {4}},
\textcircled{\fontsize{0.22cm}{0.25cm}\selectfont {17}}\}$
are not shown here.}
\label{fig:ceg18}
\end{figure}

The key observation of Kochen and Specker was that there are sets of vectors
with given exclusivity relations, where such an NCHV assignment can not be made,
these sets are then called KS sets. In the original work, a set of 117 vectors 
was considered \cite{kochen1967problem}, a simpler set was derived by Cabello 
and coworkers \cite{cabello1996bell} (see Table~\ref{tab:context} and 
Fig.~\ref{fig:ceg18}). The CEG set $\{\ket{\psi_i}\}_{i=1}^{18}$ 
forms  $9$ complete contexts, as nine orthogonal bases can be found. Consequently, any NCHV 
model should assign to one and only one vector the value $1$ in each context, and 
the total number of $1$-assignments is $9$, an odd number. On the other hand, 
each vector appears twice which implies that the total number of $1$-assignments 
should be even. Thus, we have a logical contradiction and the CEG set is a 
special case of a KS set.


{\it Different proofs of contextuality.---}
In general, we call a set of vectors  $\{\ket{\psi_i}\}$ 
in $d$-dimensional space a {\it Kochen-Specker set (KS set)} if 
there  is no $0/1$-assignment 
(denoted by $\vec{v}$) that satifies the following two conditions:
\\
(a) Two mutually exclusive events cannot both have the value $1$. This means
that $v_i v_j = 0$ if  $\ket{\psi_i}$ and $\ket{\psi_i}$ are orthogonal.
\\ 
(b) In a complete context exactly one assignment has the value $1$.
This means that $\sum_{i \in C} v_i = 1$, if $\{\ket{\psi_i}|i\in C\}$ is a set 
of mutually orthogonal vectors spanning the whole space, $\sum_{i\in C} \ketbra{\psi_i} = \id$.

The proof of quantum contextuality with a KS set is based on the
structure of quantum measurements and does not rely on any quantum
state.  For Hardy-type and GHZ-type proofs, however, also the 
predictions for some quantum states become important. 
For a given set of events with specific exclusivity relations, 
we denote by $\{C_k\}_{k=0}^K$ a subset of all contexts and by 
$\vec{p}$ a probability assignment to all events (coming from a
classical model or quantum theory).  We also write 
$\vec{p}|_{C} := \sum_{i\in C} p_i$. Then, if for an NCHV model,
\begin{equation}
\vec{p}|_{C_k} = 1, \forall k=1,2,\ldots,K \implies \vec{p}|_{C_0} = 0,
\label{eq-hardy1}
\end{equation}
while, under the same conditions, $\vec{p}|_{C_0}$ can be non-zero 
in the quantum case, one has a {\it Hardy-type proof of contextuality.} 
If one can reach $\vec{p}|_{C_0}=1$ in the quantum case, the Hardy-type 
proof is called a {\it GHZ-type proof.}

Note that in this definition $C_k$'s do not need to be complete 
contexts. In the Bell scenario, the conditions for  Hardy-type proofs
are often formulated as $\vec{p}|_{C_k}=0$ \cite{Abramsky_2011}. This is no
real difference, however. Since each context $C_k$ can be 
embedded in a complete context  $\widetilde{C}_k$, these conditions 
are equivalent to $\vec{p}_{\widetilde{C}_k \setminus C_k} = 1$. 
The latter form of conditions are more suitable for contextuality, 
where we do not embed  any context in a complete one. Below, we will also
recover the original Hardy-type proof  in the form of Eq.~(\ref{eq-hardy1}).

{\it GHZ-type proofs from KS sets.---}
The key observation for our approach is that {\it any KS proof 
can be converted to a GHZ-type proof with less events.} Similar ideas has
also been presented in Ref.~\cite{cabello1996bell}.
Let us explain this procedure with the CEG set as an example, 
the detailed discussion is given Appendix A.

If we take $\rho = |\psi_0\rangle\langle \psi_0|$ as the quantum state, 
then the probabilities $p_i = |\braket{\psi_i}{\psi_0}|^2$ for the events are
\begin{equation}
 p_0 = 1,\ p_1 = p_2 = p_3 = p_9 = p_{15} = p_{16} = p_{17} = 0,
\end{equation}
where $p_9 =0$ is due to the incomplete context 
$\{\textcircled{\fontsize{0.22cm}{0.25cm}\selectfont {0}},
\textcircled{\fontsize{0.22cm}{0.25cm}\selectfont {9}}\}$,
and
\begin{align}
  & p_4 + p_5 + p_6 = 1,\ &  p_{10} + p_{11} + p_{12} =1,\nonumber\\ 
  & p_6 + p_7 + p_8 = 1,\ & p_{12} + p_{13} + p_{14} =1,\nonumber\\
  & p_5 + p_7 + p_{14} = 1,\ & p_4 + p_{11} + p_{13} =1,
  \label{eq:ceg10c}
\end{align}
If an NCHV model satisfies Eq.~\eqref{eq:ceg10c}, then the deterministic 
probability assignment with $p_k \in\{0,1\}$ for any given hidden variable 
should also satisfy it. Summing over the equations, one has
\begin{equation*}
  (p_8 + p_{10}) = 6 - 2(p_4 + p_5 + p_6 + p_7 + p_{11} + p_{12} + p_{13} + p_{14}),
\end{equation*} 
which is an even number. Since 
$p_8 + p_{10} \le 1$ by exclusivity, $p_8 + p_{10} =0$ must hold
for any hidden variable. In the quantum case, however, 
$p_8 + p_{10} =1$, as can be directly calculated for the state
$\rho = \ket{\psi_0}\bra{\psi_0}$. Thus, a GHZ-type proof
has been constructed from CEG set. This GHZ-type proof 
consists of  $10$ events and $7$ contexts, these are shown 
in Fig.~\ref{fig:ceg10}.

{\it Proof of optimality.---}
Our method of proving optimality makes use of the graph-theoretic
approach to contextuality~\cite{cabello2014graph}, where any contextuality scenario
corresponds to a graph. More precisely, for a given set of events 
$\{e_i\}_{i\in V}$, their exclusivity relations can be represented 
by an {\it exclusivity graph} $G$. This consists of vertices, where
two vertices $i,j$ are connected (also written as $i\ue j$)if and 
only if $e_i, e_j$ are exclusive events in $V$. For example, the 
exclusivity graph of events in the GHZ-type proof in Fig.~\ref{fig:ceg10} 
is given in Appendix B. In this way, a probability assignment 
for events is automatically a value assignment for vertices in the 
exclusivity graph.

It's known that the set of probability assignments for a set of 
events $\{e_i\}_{i\in V}$ in the NCHV model are the so-called 
stable set polytope ${\rm STAB}(G)$ \cite{knuth1994sandwich, cabello2014graph},
given by 
\begin{equation}
{\rm STAB}(G) := 
\text{conv}
\left\{ \vec{v}  \mid \vec{v}\in \{0,1\}^{|V|}, v_iv_j = 0 \text{ if } i\ue j \right\},
\end{equation}
where $|V|$ is the size of $V$ and $\text{conv}\{\cdot\}$ denotes 
the convex hull. Similarly, the set of probability assignments
in quantum mechanics is the so-called theta body ${\rm TH}(G)$, 
\begin{equation}
{\rm TH}(G) := 
\left\{ \vec{v}  \mid  v_i = (\vec{u}_{i,0})^2 /\Vert \vec{u}_i\Vert^2 , \vec{u}_i\vec{u}_j^T = 0 \text{ if }i\ue j \right\}, 
\end{equation}
where the $\vec{u}_i$ are real vectors of arbitrary dimension with 
coefficients $\vec{u}_{i,k}$ and $\Vert \vec{u}_i\Vert^2 = \vec{u}_i \vec{u}_i^T$. 
Physically, this means that any probability assignment can 
always be obtained with rank-$1$ projectors and a pure state, and both can chosen
to be real.

For an arbitrary graph, a clique is a set of pairwise connected vertices.
By definition, a context just corresponds to a clique in the exclusivity 
graph. So, Hardy-type proofs and GHZ-type proofs can be phrased in the 
language of graph theory in the following way:

\begin{figure}[t]
		\centering
		\includegraphics[width=0.26\textwidth]{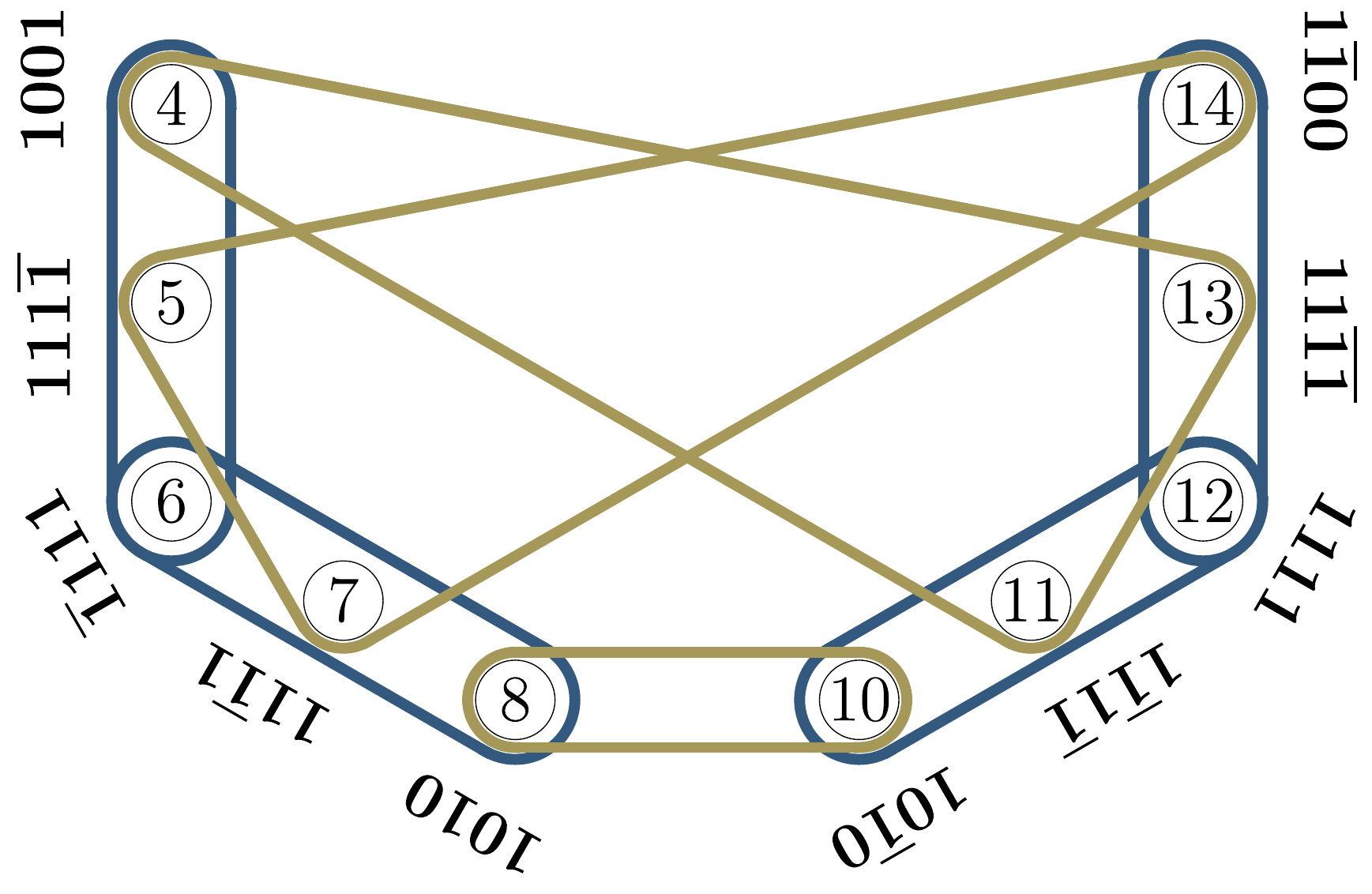}
		\caption{GHZ-type proof with $10$ events constructed from 
		CEG rays. Here, only contexts relevant for the proof are shown.}
		\label{fig:ceg10}
\end{figure}

For a given exclusivity graph, we denote by $\{C_i\}_{i=0}^k$ a 
subset of all cliques. For a given value assignment $\vec{v}$, 
we define as above $\vec{v}|_{C_i}: = \sum_{k\in C_i} v_k$. 
Then we define 
$\mathcal{V} = \left\{ v \mid v|_{C_i} = 1, \forall i=1,2,\ldots \right\}$ 
as the set of value assignments which satisfies all the conditions of 
the Hardy-type proof to be constructed, see also Eq.~(\ref{eq-hardy1}). 
If $\vec{v}|_{C_0} \equiv 0$  in the intersection of $\mathcal{V}$ and 
${\rm STAB}(G)$, but $\max \vec{v}|_{C_0} > 0$ in the intersection 
of $\mathcal{V}$ and  ${\rm TH}(G)$, then we have a Hardy-type proof. 
If  $\max \vec{v}|_{C_0} = 1$ in the latter case, then we have a 
GHZ-type proof. 

In fact, one only needs linear programming and semi-definite programming 
to check whether or not $\{C_i\}_{i=0}^k$ in a given graph provides a 
Hardy-type or GHZ-type proof.  By exhausting all the $288266$ graphs 
with less than $10$ vertices, we find no GHZ-type proof \footnote{We 
used Nauty~\cite{mckay2014practical} to generate all the non-isomorphic 
graphs with up to $9$ vertices. For each graph, we used IGraph to find 
all the maximal independent sets which is necessary for calculating 
$\mathcal{H}\cap {\rm STAB}(G)$. MOSEK was used as the solver of 
semi-definite programs for the calculation in quantum case. All the 
calculations are done with Python, which taking about $300$ hours on 
a common desktop.}. More details about the 
calculations are given in Appendix C. Thus, the ten
vectors in Fig.~\ref{fig:ceg10} constitute the minimal GHZ-type 
proof of contextuality.

We explain details of our approach using as example the exclusivity 
graph coming from the original Hardy proof in a bipartite Bell scenario. 
Let us recall the Hardy argument. If we denote by $A_i$ and  $B_j$ measurements 
with two outputs $\{0,1\} $ for Alice and Bob, the original Hardy proof 
can be phrased as
\begin{align}
\label{eq:hardyo}
p(A_1\geq B_1) & = p(B_1\geq A_2) = p(A_2\geq B_2)=1,
\nonumber\\ 
&\implies p(A_1 < B_2) \overset{\rm LHV}{=} 0,
\end{align}
where LHV stands for local hidden variable model. But, under the same conditions, 
$\max p(A_1<B_2) = (5\sqrt{5} -11) /2 \approx 0.09$ can be achieved for entangled 
quantum states. Traditionally, the conditions in Eq.~\eqref{eq:hardyo} are 
formulated as $p(A_1=0, B_1=1)=p(A_1<B_1) = 0$ etc., which is equivalent to our 
notation. There are $10$ events in the Hardy proof, which can be written as,
\begin{align}
\label{eq:chshevents}
&  [0,0|1,1], [1,0|1,1], [1,1|1,1]; \nonumber \\
&  [0,0|2,1],[0,1|2,1], [1,1|2,1];\nonumber\\
&  [0,0|2,2], [1,0|2,2], [1,1|2,2]; \text{ and } [0,1|1,2],
\end{align}
where $[a,b|i,j]$ represents the events that the outcomes are  $a,b$ 
for the measurements  $A_i, B_j$. The exclusivity relations of 
these $10$ events can be represented by the graph $G_{\rm Hardy}$ as 
in Fig.~\ref{fig:chsh}, see also Fig.~\ref{fig:hardye} in Appendix B.

In our algorithm, the representation in Eq.~(\ref{eq:chshevents}) does
not occur. Instead we start from $G_{\rm Hardy}$ as a graph and ask 
whether Hardy-type arguments can be derived from it. We only need to 
consider Hardy-type proofs with all vertices included, so we only consider 
sequences of cliques $\{C_i\}_{i=0}^k$ such that $\cup_{i=0}^k C_i = V$.
For each possible set of contexts, we run the calculations described in
Appendix C. It turns out that there are three 
Hardy-type proofs for the graph $G_{\rm Hardy}$, belonging to the 
context sets
\begin{align}
\{C_i\}_{i=0}^3  &=\{(9),(0, 1, 2), (3, 4, 5), (6, 7, 8)\},\label{eq:hardyc}\\ 
\{C_i\}_{i=0}^3  &=\{(9), (0, 1, 2), (5, 6), (3, 4, 7, 8)\},\\ 
\{C_i\}_{i=0}^4  &=\{(9),(0, 1, 4, 5), (2, 3), (5, 6), (3, 4, 7, 8)\}.
\end{align}
In the quantum case, $0.11111 \le \max (p_9) \le 0.11112$ for all three sets. 
The three sets are not entirely independent, using that the sum of probabilities 
for each clique is bounded by one, one can show their equivalence.

Representing the exclusivity graph $G_{\rm Hardy}$ by the $10$ events in 
Eq.~\eqref{eq:chshevents} amounts to making the additional assumption of 
a bipartite structure. In this case, Eq.~\eqref{eq:hardyc} recovers the 
original Hardy proof and its quantum bound $(5\sqrt{5} -11) /2 \approx 0.09017$ 
can be found with the NPA hierarchy~\cite{navascues2008convergent}.

\begin{figure}[t]
  \centering
  \includegraphics[width=0.4\linewidth]{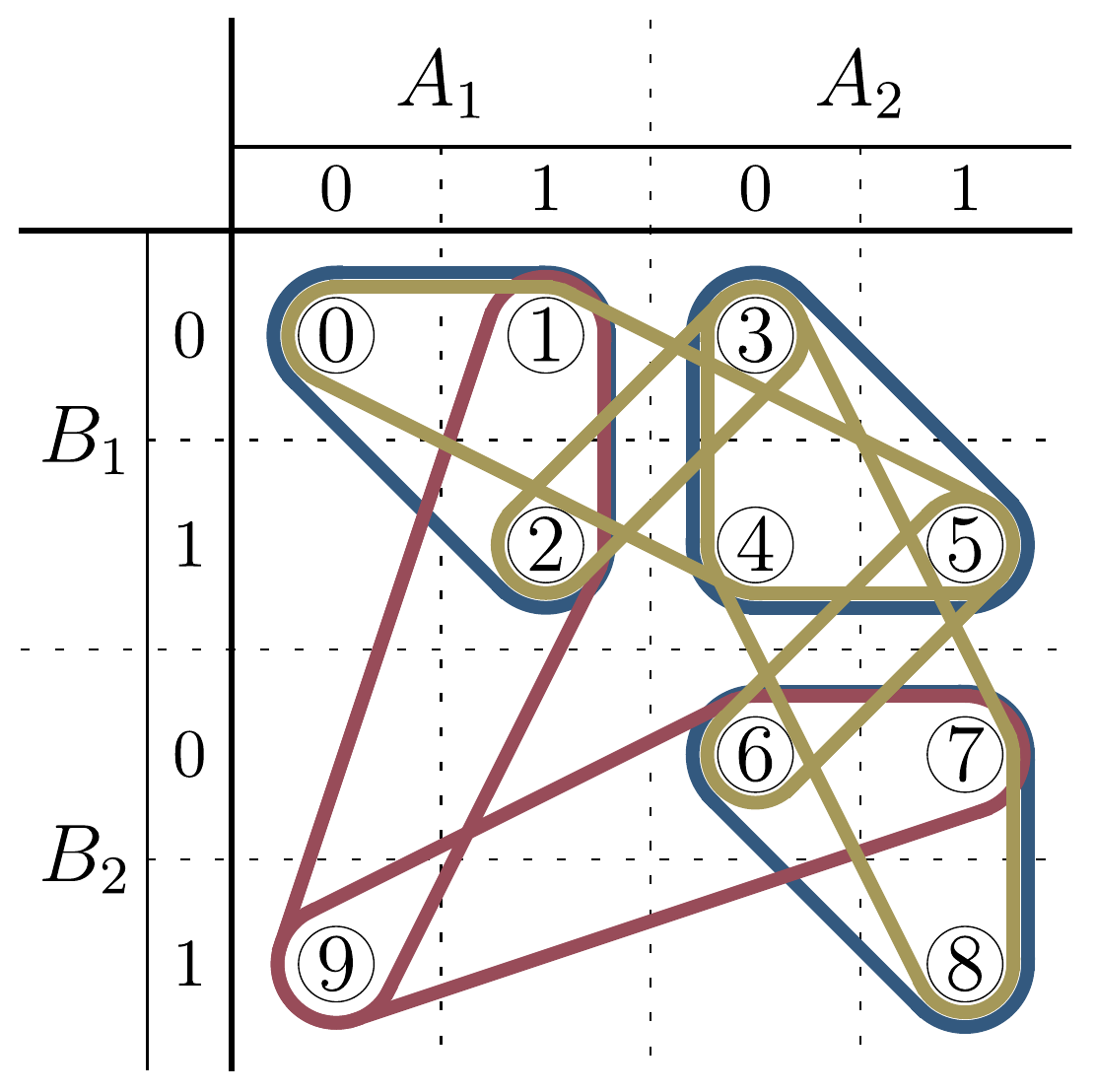}
  \caption{The contexts the exclusivity graph $G_{\rm Hardy}$, 
  related with the original Hardy proof. For each measurement 
  $A_i$'s  or $B_j$'s in the original Hardy proof, the outputs 
  are  $0$ or $1$. A vertex in the graph corresponds to the 
  joint event in its row and its column. For example, vertex  
  $2$ represents the event  $[1,1|1,1]$ originally. See also 
  Appendix B for a different representation.}
  \label{fig:chsh}
\end{figure}

{\it The proof of the Peres conjecture.---}
Based on the fact the size of the minimal GHZ-type proof 
is $10$, we prove that the size of the minimal
KS set is  $18$. The key idea is that to show that any
KS set with 17 vectors would result in a GHZ-type proof
with 9 events. First, we need one technical Lemma.

{\bf Lemma.}
{\it For a given KS set of vectors, if each vector is 
contained in exactly one complete context, then
there should be at least four complete contexts in
the whole set.}

The idea of the proof is that if there are three or less
complete contexts, then a classical assignment can be directly
written down, details are given in Appendix D.
Now we can prove the final result of this paper:

{\bf Theorem [Peres conjecture].}
{\it The size of the minimal Kochen-Specker vector set is $18$. 
So, the construction by Cabello, Estebaranz, and Garc\'{i}a-Alcaine
is optimal.}

It was proven already that in three-dimensional space, any  KS 
set should contain no less than $22$ vectors, and for $d=4$ it 
should contain no less than $18$ vectors~\cite{uijlen2016kochen, 
pavicic2005kochen}. Here we only need to consider the dimensions $d\geq 5$. 
The Lemma implies that we only need to consider the case where there 
are at least two contexts sharing at least one common vector. Because 
otherwise, we would have four complete contexts, meaning that the size 
of the KS set is at least $4d\geq 20$. 

Let us say that the $d$-dimensional minimal KS set contains $n$ vectors 
and there are two complete contexts $C_1, C_2$ with non-empty intersection
and $\ket{\psi_0}\in C_1 \cap C_2.$ Since $C_1$ must contain at least two 
vectors which are not in  $C_2$, we have $|C_1 \cap C_2| \leq d-2$. This 
implies that $|C_1 \cup C_2| \geq d+2$. Now, as above (see also 
Appendix A) we can assume that a system is in the quantum
state $\ket{\psi_0}$, remove the vectors in $|C_1 \cup C_2|$ and arrive
at a GHZ-type proof. This GHZ-type proof would have no more than $n-(d+2)$ 
vectors by construction, but as we know, it must contain at least 10 vectors. This proves that 
$n\geq 18$ if $d\geq 6$. For the case $d=5$, some extra considerations
are needed, details are given in Appendix D.
So, the Peres conjecture is proved.

{\it Conclusion and Discussion.---}
By proposing a systematical approach to find Hardy-type
and GHZ-type proofs, we showed that the minimal size of 
a GHZ-type proof of contextuality is $10$. Based on this, 
we proved the Peres conjecture for contextuality, stating
that the minimal KS set consist of $18$ vectors. There
are several directions to extend our results. First, it 
would be interesting to characterize the minimal GHZ-type 
proofs in fixed dimensions, e.g., for $d=3$. Here the techniques 
of \cite{zhenpeng2020,jia2020,miguelbounding}, may be useful. Furthermore, 
while we considered only KS sets of vectors (or one-dimensional
projectors), one may define KS sets also using projectors of 
higher rank. 

From the viewpoint of quantum information theory, it would be
highly desirable to study the usefulness of the minimal KS 
scenarios for information processing~\cite{cubitt2010improving,cabello2011hybrid,aolita2012fully}. For instance, given the 
quantum state $\ket{\psi_0}$ and the measurements from Fig.~\ref{fig:ceg10}, 
are there any computations that can be carried out better than in a
classical model? Similar questions have been discussed for
measurement-based quantum computation  
\cite{Raussendorf_2013, anders2009computational}, and the answer may shed light on
the role of contextuality in quantum computing.

\begin{acknowledgments}
We thank Ad\'an Cabello and Matthias Kleinmann for discussions. This work was supported by 
the DFG, the ERC (Consolidator Grant No. 683107/TempoQ) and the
Humboldt foundation.
\end{acknowledgments}

\appendix

\section{From KS sets to GHZ-type proofs}
\label{sec:ks2ghz}
In this section we will explain in detail how any KS set of vectors
can be converted into a GHZ-type proof. Let us recall the strategy. 
Starting from an KS set (as in Fig.~\ref{fig:ceg18}) we take one
arbitrary vector (here, the vector $\ket{\psi_0}$) from the set, and
assume this as a quantum state. Then, we remove the vector
$\ket{\psi_0}$ and all the vectors orthogonal to it from the set. 
This, of course, transforms some of the complete contexts in the
original KS set into incomplete ones, as also happens in Fig.~\ref{fig:ceg10}. 
It remains to show that the reduced graph gives rise to a GHZ-type proof. 
This was explicitely calculated for the CEG set in Eqs.~(\ref{eq:ceg10c}),
but for a general initial KS set a more general argument  is required. 

In the language of graph theory, a KS proof can be expressed as follows. 
For a given graph $G$ and a set $\{C_k\}_{k=0}^{K}$ of maximal cliques 
in $G$, we have a set of conditions for a probability assignment 
$\vec{p}$ of  $G$
\begin{align}
\label{eq:kscond}
\text{completeness: }& \vec{p}|_{C_k} = 1, \forall k=0,\ldots,K,\nonumber\\
\text{exclusivity: }& p_i p_j = 0, \text{ if } i\ue j, \forall i,j.
\end{align}
If all conditions in Eq.~\eqref{eq:kscond} cannot be fulfilled simultaneously 
in an NCHV model, while they are satisfied by the probability assignment 
$\vec{p}$ induced by $p_i := \bra{\psi_i}\rho\ket{\psi_i}$ for a given 
set $\{\ket{\psi_i}\}$ of vectors and any quantum state $\rho$, then we 
have a KS proof.

Such a KS proof is said to be tight if all the conditions can be 
satisfied simultaneously by an NCHV model after removing a single 
(but arbitrary) completeness condition. Any KS proof can always 
be transformed into a tight KS proof by removing the reluctant 
completeness conditions step by step from the original set. 
So, we can assume without loosing generality that the KS proof 
is tight.

Let us take from an NCHV model the probabilities 
$\vec{p} = \sum_{\lambda\in \Lambda} \mu(\lambda) \vec{p}_\lambda$, 
given as the convex decomposition of $\vec{p}$ into the deterministic 
probability assignments $\vec{p}_\lambda$ associated with fixed 
hidden variables, where $\mu(\lambda)>0$ and 
$\sum_{\lambda\in \Lambda} \mu(\lambda) =1$. 
Then, for any complete context
\begin{equation}
1=\vec{p}|_{C_k} =  \sum_{\lambda\in\Lambda} \mu(\lambda) \vec{p}_{\lambda}|_{C_k}
\implies \vec{p}_{\lambda}|_{C_k} = 1, \forall \lambda.
\end{equation} 
If we assume that all the exclusivity conditions are satisfied, 
then we can take for the $K+1$ complete contexts an arbitrary ordering 
$\{C_{k_i}\}_{i=0}^K$, and, since the proof is tight, the relations 
$\vec{p}|_{C_{k_i}}=1, \forall i=1,\ldots,t$ are possible in an NCHV 
model, but then it's not possible anymore to obey the condition 
$\vec{p}|_{C_{k_0}}=1$. 
In fact, 
\begin{align}
\label{eq:ksghz}
&  \vec{p}|_{C_{k_i}} = 1, \forall i=1,\ldots,K 
\implies  \vec{p}_{\lambda}|_{C_{k_i}} = 1, \forall i=1,\ldots,t\nonumber\\
&  \implies \vec{p}_\lambda|_{C_{k_0}} = 0 \implies \vec{p}|_{C_{k_0}} = 0.
\end{align}
Thus, we have a GHZ-type proof, as $\vec{p}|_{C_{k_0}} = 1$ holds for any 
quantum state. But this proof is state-independent and still has the same 
number of events as the KS set.

If we take any state $\ket{\psi_i}$ where $i\not\in C_{k_0}$ as a quantum state,
we have
\begin{equation}\label{eq:extracond}
p_i = 1, p_j = 0, \text{ if } i\ue j, \forall j.
\end{equation} 
Hence, by assuming the additional exclusivity conditions in Eq.~\eqref{eq:extracond}, 
we can simplify the GHZ-type proof in Eq.~\eqref{eq:ksghz}: Events related 
with vertices $\{i\} \cup \{j|i\ue j, \forall j\}$ can be removed from 
the GHZ-type proof in Eq.~\eqref{eq:ksghz}. This implies that all vectors
in the contexts where $\ket{\psi_i}$ belongs to, disappear. Other contexts
${C_{k_\ell}}$ with $\ell=0,\ldots,K$ are not complete anymore, but this does
not change the logic from Eq.~\eqref{eq:ksghz}.  So we arrive at the desired
state-dependent GHZ-type proof. Finally, since the choice of the set 
$C_{k_0}$ is arbitrary, the choice of $\ket{\psi_i}$ also is.

\section{Examples of exclusivity graphs}
\label{sec:exghs}
\begin{figure}[t!]
		\centering
		\includegraphics[width=0.35\textwidth]{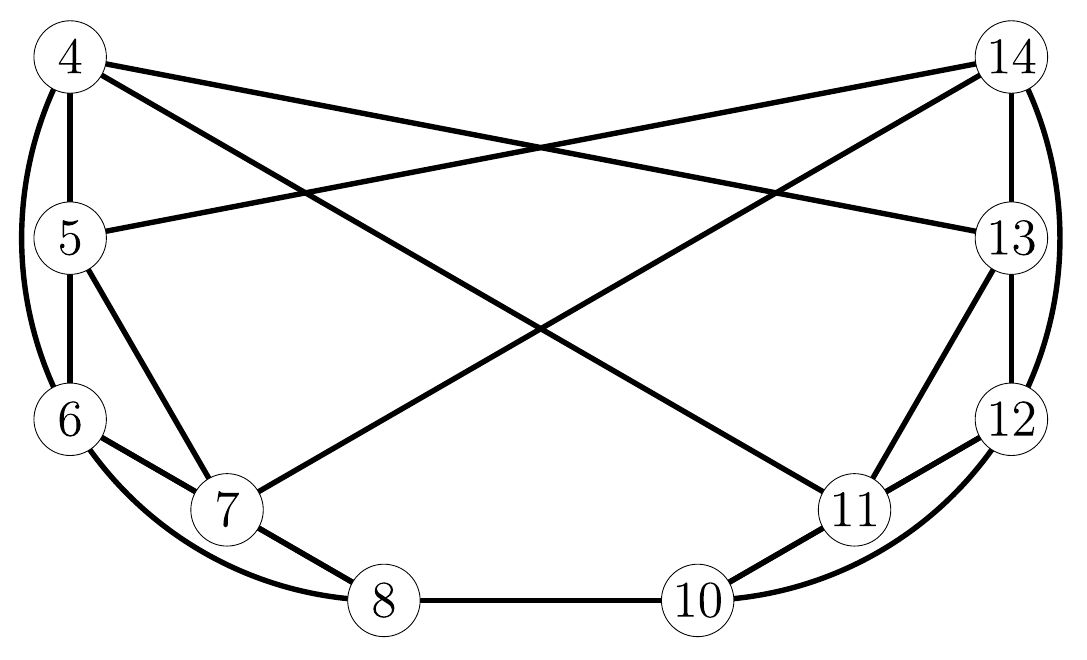}
		\caption{Exclusivity graph of events used in the GHZ-type proof in Fig.~\ref{fig:ceg10}.}
		\label{fig:ceg10e}
\end{figure}

\begin{figure}[t!]
		\centering
		\includegraphics[width=0.22\textwidth]{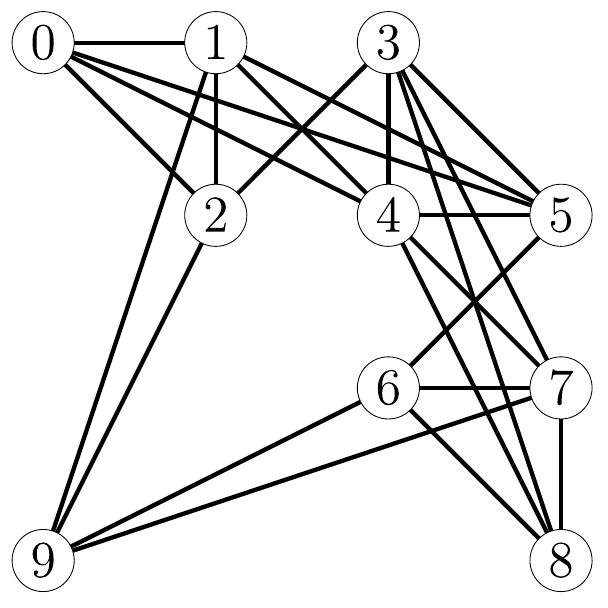}
		\caption{Exclusivity graph of events used in the original Hardy-type proof, 
		see Fig.~\ref{fig:chsh}.}
		\label{fig:hardye}
\end{figure}

In this section we present in detail the exclusivity graphs for two
important examples in the main text. Fig.~\ref{fig:ceg10e} shows the
exclusivity graph of events used in the GHZ-type proof in 
Fig.~\ref{fig:ceg10} in the main text. Fig.~\ref{fig:hardye} presents the
exclusivity graph of events used in the original Hardy-typy 
proof, see Fig.~\ref{fig:chsh} in the main text.


\section{Detailed explanation of steps}
\label{sec:steps}
In this section, we describe in detail the calculation of the classical 
maximum as well as the quantum maximum for probability distributions 
obeying the conditions of the Hardy-type proof, see also Eq.~\ref{eq-hardy1}. 
The probability disributions obeying these constraints were denoted by $\mathcal{V}$
in the main text. So, for the classical case of NCHV models, one has to consider
probability distributions in $\mathcal{V} \cap {\rm STAB}(G)$, and for
the quantum case, probability distributions in $\mathcal{V}\cap {\rm TH}(G).$

\subsection{NCHV models: $\mathcal{V} \cap {\rm STAB}(G)$}
For a given graph $G(V,E)$, we denote the character function 
for a subset $S$ of vertices  as
\begin{equation}
\mathds{1}_{S} := \{\vec{v}|v_i = \delta(i\in S)\},
\end{equation} 
where $\delta(x) = 1$ if  $x$ is true, otherwise, $\delta(x)=0$. 

Let us denote $\mathcal{I}$  as the set of all independent sets of $G$. 
An independent set of $G$ is a set of vertices, in which any two vertices 
are not connected. An independent set is said to be maximal if it is not 
a subset of any other independent set. By definition, 
\begin{equation}
  {\rm STAB}(G) = {\rm conv} \{\mathds{1}_I | I \in \mathcal{I}\}. 
\end{equation} 
For any set of probabilities $\vec{v}\in {\rm STAB}(G)$ and any clique $C$, 
one has $\vec{v}|_{C} \le 1$. If we write 
$\vec{v} = \sum_{i=1}^t x_i \mathds{1}_{I_i}$ with $x_i >0$, then 
\begin{equation}
\vec{v} \in \mathcal{V} \implies \mathds{1}_{I_i} \in \mathcal{V}, \forall i=1,\ldots,t.
\end{equation} 
For two independent sets $I_1$ and  $I_2$ with $I_1 \subseteq I_2$ we have
\begin{equation}
\mathds{1}_{I_1} \in \mathcal{V} \implies \mathds{1}_{I_2} \in \mathcal{V}.
\end{equation} 
By definition, $\mathds{1}_{I} \in \mathcal{V} \cap {\rm STAB}(G)$ means that 
the independent set $I$ has non-empty intersection with each $C_i$ for  
$i=1,\ldots,k$. On the other hand, $\mathds{1}_{I}|_{C_0} = 0$ is equivalent 
to  $I \cap C_0 = \emptyset$. Hence, to check whether  $\vec{v}|_{C_0} \equiv 0$ 
for any $\vec{v}$ in the intersection of $\mathcal{V} \cap {\rm STAB}(G)$, we 
can just calculate 
\begin{equation}
  \mathcal{I}_{\rm pre} :=  \{ I | I\cap C_i \neq \emptyset, \forall i=1,\ldots,k\} \cap \mathcal{I}_{\max},
\end{equation}
where $\mathcal{I}_{\max}$ is the set of all maximal independent 
sets. Then we check whether the sets $\mathcal{I}_{\rm pre}$ and  
$\cup_{I\in \mathcal{I}_{\rm pre}} I \cap C_0$ are empty or not. 

If $\cup_{I\in \mathcal{I}_{\rm pre}} I \cap C_0 = \emptyset$, we 
have  $\vec{v}|_{C_0} \equiv 0$, $\forall \vec{v} \in \mathcal{V}\cap {\rm STAB}(G)$. 
If $\mathcal{I}_{\rm pre}$ is empty, then it implies that we already have 
an candidate of Hardy-type proof by choosing one $C_i$ as the 
new $C_0$ while keeping the rest $C_i$'s as conditions.

\subsection{Quantum mechanics: $\mathcal{V}\cap {\rm TH}(G)$}
In quantum theory, we have to calculate
\begin{align}
  \max : \; &\vec{v}|_{C_0} \nonumber \\
  \text{subject to} : \; & \vec{v}|_{C_i} = 1, \forall i=1,\ldots,k \nonumber \\
                    & \vec{v}\in {\rm TH}(G)
\end{align}
For a set of probabilities $\vec{v}\in {\rm TH}(G)$ in the theta 
body we have by definition $v_i = (\vec{u}_{i,0})^2/\Vert \vec{u}_i\Vert^2 $ where 
$\vec{u}_i \vec{u}_j^T = 0$ if $i\ue j$. This implies that
\begin{equation}\label{eq:mat}
  T:= \begin{bmatrix} 1 & \vec{v}\\ \vec{v}^T & A \end{bmatrix} \succeq 0 
\end{equation} 
where the entries of the matrix $A = (A_{ij})$ 
are given by
$A_{ij} = (\vec{s}\vec{u}^T_i \vec{u}_i \vec{u}_j^T \vec{u}_j \vec{s}^T) 
/ 
\Vert \vec{u}_i\Vert^2 \Vert \vec{u}_j\Vert^2$ and $\vec{s} = (1,0,\ldots,0)$. 
Hence, $A_{ii} = v_i$ and $A_{ij} =0$ if $i\ue j$. The positivity of
$T$ follows from the fact $T = P^T P$, where the columns of $P$ are
\begin{equation}
		P = [\vec{s}^T\ (\vec{u}_1^T \vec{u}_1 \vec{s}^T) /\Vert \vec{u}_1 \Vert^2 \ldots (\vec{u}_n^T \vec{u}_n \vec{s}^T) /\Vert \vec{u}_n \Vert^2].
\end{equation}

Conversely, consider that we have a given positive semidefinite matrix 
$T$ as in Eq.~\eqref{eq:mat}, where $diag(A) = \vec{v}$ and $A_{ij} = 0$ 
if  $i\ue j$. Then, using the Cholesky decomposition, $T$ can always be 
decomposed as  $ T = P^T P$ where the columns of $P$ can be written
as $P = [\vec{s}^T\ \vec{\mu}_1^T\ \ldots\ \vec{\mu}_n^T]$. The fact 
that $diag(A) = \vec{v}$ and the structure of the matrix in Eq.~\eqref{eq:mat}
implies that $v_i = \Vert\vec{\mu}_i\Vert^2 = \vec{\mu}_i \vec{s}^T.$
So we have $v_i =(\vec{\mu}_i\vec{s}^T)^2 /\Vert{\mu}_i\Vert^2$. If $i\ue j$, then  $A_{ij} =0 = \vec{\mu}_i \vec{\mu}_j^T$. Hence, $\vec{v}\in {\rm TH}(G)$
if the quantum state is described by the vector $\vec{s}.$

Thus, the condition $\vec{v} \in {\rm TH}(G)$ is equivalent to 
\begin{equation}
  \begin{bmatrix} 1 & \vec{v} \\ \vec{v}^T & A \end{bmatrix} \succeq 0,\ diag(A) = \vec{v}, \ A_{ij}=0, \text{ if } i\ue j.
\end{equation} 
The whole semi-definite program for the quantum case is
\begin{align}\label{eq:sdp}
  \max : \; &\vec{v}|_{C_0}\nonumber\\
  \text{subject to }: \; & \vec{v}|_{C_i} = 1, \forall i=1,\ldots,k,\nonumber\\
				    & \begin{bmatrix} 1 & \vec{v} \\ \vec{v}^T & A \end{bmatrix} \succeq 0,\nonumber\\
					& diag(A) = \vec{v},\nonumber\\ 
					& A_{ij}=0, \text{ if } i\ue j.
\end{align}

\section{Proofs of the Lemma and the Theorem}
\label{sec:proofs}
In this section, we give the detailed proofs of the Lemma and the Theorem
in the main text. We repeat the statements here for better readability.

{\bf Lemma.}
{\it For a given KS set of vectors, if each vector is 
contained in exactly one complete context, then
there should be at least four complete contexts in
the whole set.}

{\it Proof.}
Assume that there are only three different complete contexts 
$C_1, C_2, C_3$ without any intersection. We can always 
pick $\ket{\psi_1} \in C_1 $ and $\ket{\psi_2} \in C_2$ such that 
$\braket{\psi_1}{\psi_2} \neq 0$. Let us denote by
$C_3^i := \{\ket{\psi} \;|\; \ket{\psi} \in C_3, \braket{\psi}{\psi_i} \neq 0\}$
for $i=1,2$ the vectors in $C_3$ which are not orthogonal
to $\ket{\psi_1}$ or $\ket{\psi_2}$. The sets $C_3^i$ each contain at least
two vectors, and the subspace corresponding to $\ket{\psi_i}$ is 
included in the one spanned by vectors in $C_3^i$ for $i=1,2$. 
Since $\braket{\psi_1}{\psi_2} \neq 0$ we must have $C_3^1 \cap C_3^2 \neq \emptyset$
and we take a vector $\ket{\psi_3}$ in it. By assigning $1$'s to 
$\ket{\psi_1}, \ket{\psi_2}, \ket{\psi_3}$, and $0$'s to all the remaining vectors, 
we have an valid $0$-$1$ assignment. So, there is no KS proof 
in this case.
\qed

{\bf Theorem [Peres conjecture].}
{\it The size of the minimal Kochen-Specker vector set is $18$. 
So, the construction by Cabello, Estebaranz, and Garc\'{i}a-Alcaine
is optimal.}

{\it Proof.}
It was proven already that in three-dimensional space, any  KS 
set should contain no less than $22$ vectors, and for $d=4$ it 
should contain no less than $18$ vectors~\cite{uijlen2016kochen, 
pavicic2005kochen}. Here we only need to consider the dimensions $d\geq 5$. 
The Lemma implies that we only need to consider the case where there 
are at least two contexts sharing at least one common vector. Because 
otherwise, we would have four complete contexts, meaning that the size 
of the KS set is at least $4d\geq 20$. 

Let us say that the $d$-dimensional minimal KS set contains $n$ vectors 
and there are two complete contexts $C_1, C_2$ with non-empty intersection
and $\ket{\psi_0}\in C_1 \cap C_2.$ Since $C_1$ must contain at least two 
vectors which are not in  $C_2$, we have $|C_1 \cap C_2| \leq d-2$. This 
implies that $|C_1 \cup C_2| \geq d+2$. Now, as above (see also 
Appendix~\ref{sec:ks2ghz}) we can assume that a system is in the quantum
state $\ket{\psi_0}$, remove the vectors in $|C_1 \cup C_2|$ and arrive
at a GHZ-type proof. This GHZ-type proof would have no more than $n-(d+2)$ 
vectors by construction, but as we know, it must contain at least 10 
vectors. This proves that $n\geq 18$ if $d\geq 6$.

It remains to consider the dimension $d=5$ and here two conditions 
on possible KS sets with less than 18 vectors can be derived. First, 
if there are two complete contexts $C_1, C_2$ in the KS set such 
that they share one or two vectors, then we have that $|C_1 \cup C_2| \ge 8$. 
This implies, as before, $n\ge 8+10 = 18$. Since in $d=5$ two different 
contexts have maximally three vectors in common, it follows that for 
all $i,j$ we have $|C_i\cap C_j| \in \{0,3\}$ (first condition). 

Second, consider the case that a vector is in two complete
contexts, $\ket{\psi_0} \in C_1 \cap C_2$ and there is a vector
$\ket{\psi_1}$ in the KS set that is orthogonal to $\ket{\psi_0}$, but 
$\ket{\psi_1} \not \in C_1 \cup C_2$. Then we may choose
$\ket{\psi_0}$ as a quantum state and remove the 7 vectors in $C_1 \cup C_2$
as well as $\ket{\psi_1}$ and arrive at a GHZ-type proof. So also
in this case $n \geq 18$, and it follows that a vector like  $\ket{\psi_1}$
cannot exist (second condition).

We now show that there must be three overlapping contexts. If this 
were not the case, we would have a given complete context, say $C_1$, 
such that there is at most one other complete context $C_2$ that has non-empty 
intersection with it, and $C_2$ has no intersection with further complete 
contexts. If there is a feasible $0$-$1$ value assignment to the vectors 
which are not in $C_1\cup C_2$, then we can just assign the value $1$ to 
one vector $\ket{\psi_0}$ in  $C_1 \cap C_2$ and  $0$ to the remaining 
vectors in  $C_1 \cup C_2$. This will result in a valid {\it global} $0$-$1$ 
value assignment, satisfying the completeness relations and the exclusivity 
relations for the original KS set, due to the second condition.
But such a global assignment is not possible, as we have a KS set.
It follows that already an assignment for the vectors  which are not 
in $C_1\cup C_2$ is not possible, this implies that we can have a simplified 
KS set by removing the vectors in $C_1 \cup C_2$. But this in contradiction
to the optimality of the KS set.

So, we are left with the situation that there is a complete context $C_1$ 
which has intersection with two different complete contexts $C_2$ and $C_3$. 
Since $|C_1 \cap C_2| = |C_1 \cap C_3| = 3$ we must have 
$|C_1 \cap C_2 \cap C_3| \geq 1.$ So we can take 
$\ket{\psi_0} \in C_1 \cap C_2 \cap C_3.$  Let us consider the set 
$C_1 \cup C_2 \cup C_3.$ If this consists of eight or more vectors, 
we can take as usual $\ket{\psi_0}$ as a quantum state and remove 
the vectors in $C_1 \cup C_2\cup C_3$, arriving at a GHZ-type proof. So, 
in this case we have $n \geq 18$.

It may be, however, that $C_1 \cup C_2 \cup C_3$ consists of seven vectors 
only. Then, up to some permutations, we can assume that these vectors are
given by
$\mathcal{I}=\{\ket{\psi_0}, \dots , \ket{\psi_6}\}$ and we have
$C_1 = \{\ket{\psi_0}, \ket{\psi_1},\ket{\psi_2},\ket{\psi_3},\ket{\psi_4}\}$,
$C_2 = \{\ket{\psi_0}, \ket{\psi_1},\ket{\psi_2},\ket{\psi_5},\ket{\psi_6}\}$,
and
$C_3 = \{\ket{\psi_0}, \ket{\psi_3},\ket{\psi_4},\ket{\psi_5},\ket{\psi_6}\}$.
Now, there must be a fourth context $C_4$, containing a vector from the set $\mathcal{I}$.
Otherwise, due to the second condition, any proper $0$-$1$ assignment to
the vectors outside of $\mathcal{I}$ can be extended to an assignment of the full KS set
(by assigning $1$ to $\ket{\psi_0}$  and $0$ to the rest in $\mathcal{I}$), and the KS set 
cannot be minimal. 

Since $C_4$ contains a vector from $\mathcal{I}$, it has overlap with two $C_i$, so let us 
assume that $C_4 \cap C_1 \neq \emptyset \neq C_4 \cap C_2$. But then, considering
the contexts $C_1, C_2, C_4$, an argument as above shows that either $n \geq 18$ or
$|C_1 \cup C_2 \cup C_4| =7,$ implying that $C_4 \subset \mathcal{I}$. But the latter is 
not possible, one cannot find four different complete contexts in the seven element set 
$\mathcal{I}$ such that the first condition holds. So, the Peres conjecture is proved.
\qed


\bibliographystyle{apsrev4-1}
\bibliography{hardyghzks.bib}

\end{document}